# Scaling and crossover phenomena in anomalous helium sequence

N. K. Das[a*], R.K. Bhandari[a], P. Sen[b], and B. Sinha[a, b]

[a] Variable Energy Cyclotron Centre, 1/AF Bidhan Nagar, Kolkata-700064, India
Email : nkdas@veccal.ernet.in

[b] Saha Institute of Nuclear Physics, 1/AF Bidhan Nagar, Kolkata-700064, India
Email: prasanta.sen@saha.ac.in

**Abstract**

Anomalous temporal fluctuations of helium concentrations in spring emanations have been observed on a number of occasions prior to some major seismic events. Several recent studies have shown that a wide variety of natural systems display significant fluctuations that may be characterized by long-range power-law correlations. We have applied detrended fluctuation analysis (DFA) to characterize preseismic helium anomalies and to probe the relationship between two classes of apparently irregular helium sequences. Application of the DFA technique reveals a crossover phenomenon that distinguishes short-range from long-range scaling exponents; the crossover corresponds to a transition from nonpersistent to persistent traits in the helium time series. Our findings imply a significant statistical correlation between anomalous helium concentration and a fluctuation exponent. This analytical approach appears to be a promising way for identifying anomalous helium fluctuations as signals precursory to an earthquake.

*Head, Helium Laboratory, Variable Energy Cyclotron Centre, 1/AF, Bidhannagar, Kolkata-700064, West Bengal, India, E-Mail: nkdas@veccal.ernet.in,
Tel: +91-33-23184105, Fax: +91-33-23346871

# 1. Introduction

A solution to the problem of earthquake prediction requires a thorough understanding of geochemical anomalies, including changes in terrestrial gas compositions that precede an earthquake. In this regard, we have been, for the past several years, performing real-time monitoring of helium and radon concentrations in gases that emanate from the thermal spring at Agnikunda (69°C), Bakreswar, W. Bengal, India (Das et al., 2006; Ghose et al., 1996). We have set up a multi-parametric geochemical laboratory that continuously assays helium at 60-minute intervals throughout the year. In the present study, we analyze helium concentrations taken over a three-month period from February through April 2007.

## 1.1 Geological Setting of the Monitored Spring

Figure 1 shows a geological map of the area surrounding the thermal spring. The thermal springs at Bakreswar (23°52′30′′N, 87°02′30′′E) are situated in an area that extends the outpouring of ancient Rajmahal Volcanism (115 ma). Geochemical analyses indicate that the spring fluids contain $Cl^-$, $F^-$, $SO_4$, B, and significant quantities of helium; these, together with a high thermal gradient in the area, imply the presence of deep-seated contributions to spring effluents. The area is marked with a 1.2 km shear zone, which is characterized by 50 m wide brecia/cherty quartzite trending in the N-S direction. The spring is located along a prominent NW-SW lineament, 25 km SW of Rajmahal Trap, and is confined to igneous intrusions on the northern periphery of the Gondwana sediments. (Nagar et al., 1996)

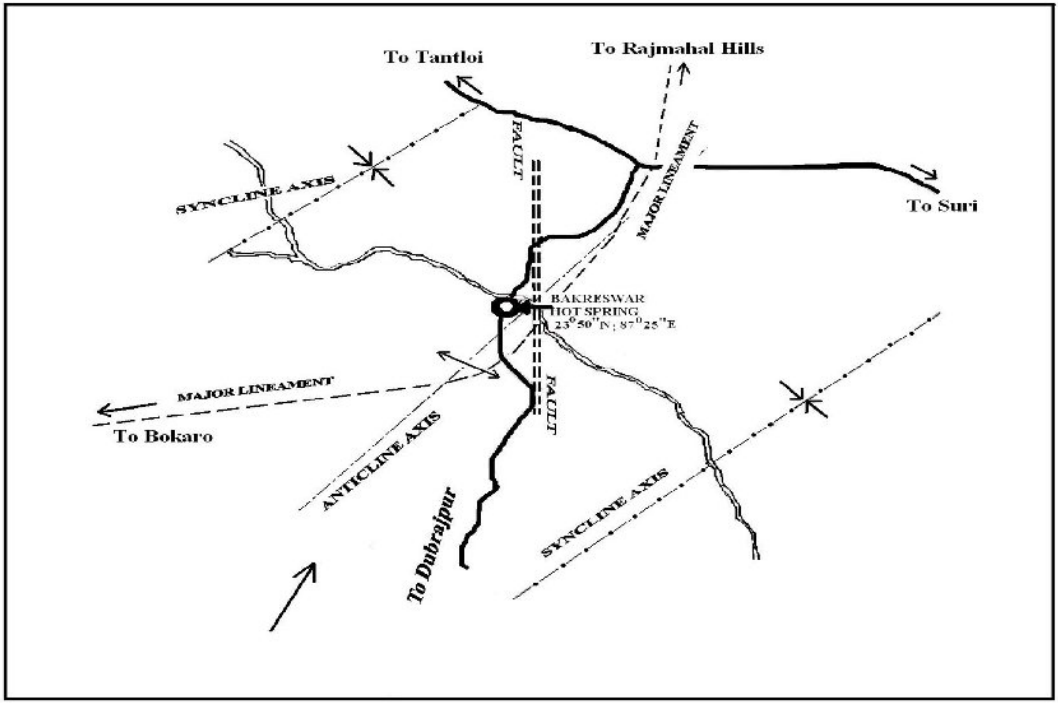

Fig. 1   Geological disposition of the thermal spring area, Bakreswar

**1.2 Detrended Fluctuation Analysis**

To determine scaling properties from the observed helium sequence, the method of detrended fluctuation analysis (DFA) has been used (Peng et al., 1994). This method has been successfully applied to diverse phenomena in which complete information about the underlying processes could not be accessed. However, several long-range power-law correlations have been observed (Mantegna and Stanley, 1995). Substantial progress has been made in such fields as DNA studies (Stanley et al., 1999), meteorology (Ivanova and Ausloos, 1999), economics (Ausloos and Ivanova, 2001), physical and biological sciences (Stanley et al., 1993), and self-organized critical systems (Bak et al., 1989). Even though trends and non-stationarities are invariably present in a time series generated by natural and physical processes, DFA analysis is capable of eliminating apparent correlations that are, in fact, artifacts arising from nonstationarities. The method provides a parameter $\alpha$, the scaling exponent, that represents correlation properties of the time series.

A scaling exponent is likely to be of value in explorations of the underlying scaling properties of the helium time series that result when underground rock masses are subjected to stress alterations. Seismically induced stresses give rise to a long-range mosaic pattern in the strain field throughout the lithosphere; such patterns constitute long-distance precursory phenomena caused by preparatory earthquake processes (Morgounov, 2004). Because of this, scaling exponents are of considerable value as quantitative signatures of dynamic processes that govern anomalous variations in real-time data. Identification of scaling behavior, in turn, provides a succinct description of system fluctuations across a range of time scales.

This paper uses the scaling exponent derived from fluctuation analysis to describe characteristics inherent in the experimentally obtained helium sequence. We also attempt to determine the effects of anomalous variations that are embedded in the sequence.

**2. Methodology**

Our helium data exhibit fluctuations within $2\sigma$ (where $\sigma$ is the standard deviation) of the average helium concentration, but there are also spikes beyond $\pm 2\sigma$. However, noise in a time sequence could mask important dynamical information, thereby affecting computational precision. To eliminate noise, a singular value decomposition technique was employed. After removing the noise, the helium sequence was organized into two time series: an anomalous series, corresponding to seismically disturbed phases, and a smoothed series, corresponding to seismically quiescent phases. The anomalous series contains the anomalous data ($x_i > 2\sigma$), while the smoothed series ($x_i \leq 2\sigma$) is the time sequence remaining after the anomalous spikes were removed.

Let the helium time series be represented by $x(k)$ of length $N$. The series was integrated after subtracting the average value of $x$ and takes the form

$$y(k) = \sum_{i=1}^{k} [x(i) - \bar{x}]$$

Here $x(i)$ is the $i^{th}$ discrete value in the finite helium series and $\bar{x}$ stands for the average helium concentration. Then this cumulative (integrated) departure from the mean was divided into windows of equal length $m$; for each window, a linear least-square line, representing the trend in the interval, was fit to the data. The $y$-coordinates of the straight-line segments were subtracted from the integrated values in each interval to detrend the series; these are denoted by $y_m(k)$. Root-mean-square fluctuations $F(m)$ around the regression line were then calculated over all intervals. The fluctuations are given by

$$F(m) = \sqrt{\frac{1}{N} \sum_{k=1}^{N} [y(k) - y_m(k)]^2}$$

This computation was repeated over all time scales (window sizes) to determine the relationship between the average fluctuation, $F(m)$, and window size $m$. A log-log plot of $F(m)$ vs $m$ was then linearly regressed to obtain the slope, $\alpha$, the scaling exponent, which is a typical fingerprint of the scaling behavior intrinsic to the data.

A linear relationship on a log-log plot indicates power-law (fractal) scaling. Here, a power law relation, $F(m) \sim m^{\alpha}$, indicates long-range memory dependence. Different values of $\alpha$ signify different levels of temporal correlation in fluctuations over different time scales. Values of $\alpha$ in the range $1.0 < \alpha < 1.5$ correspond to long-range negatively correlated (antipersistent) noise (Ivanova and Ausloos, 1999) that somewhat resembles brown noise. Values in the range $0.5 < \alpha < 1.0$ correspond to processes in which fluctuations are positively correlated (persistent noise); these show long-range power-law scaling behavior (Hoop and Peng, 2000).

## 3. Results and Discussion

Figure 2 shows the temporal variation in the original helium concentration, including anomalous peaks ($>\pm 2\sigma$). It is conjectured that there is a trade-off between distinct anomalies in helium concentration and earthquake nucleation processes that arise from crustal strain episodes.

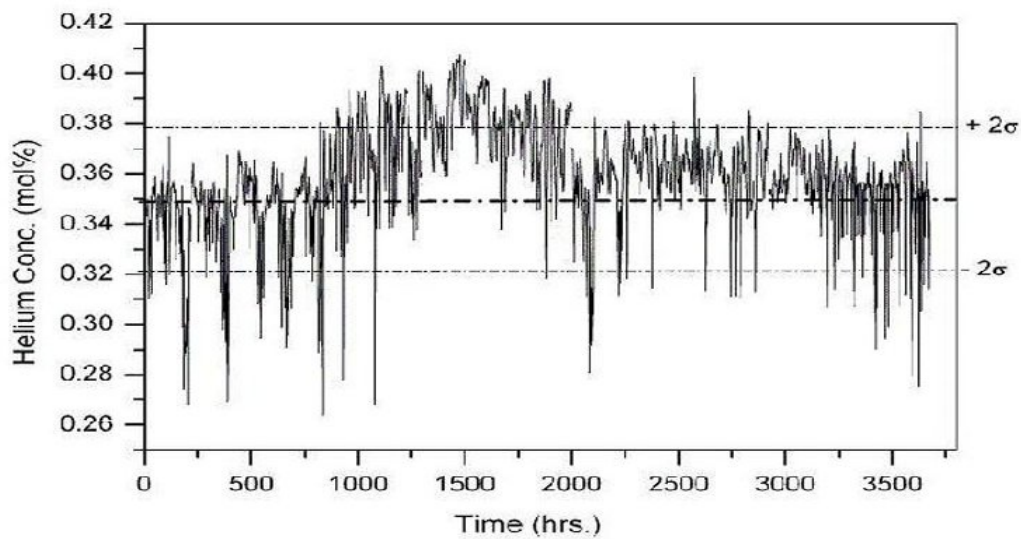

Fig. 2 Time series plot of helium concentration

The presence of trends and nonstationarities in time series derived from real world systems hinder the use of autocorrelation functions and power spectra; nevertheless, these analyses can still provide a qualitative impression of the time series. Figures 3a and 3b show that the autocorrelation functions (ACF) decay slowly to zero, with the first zero crossing of the ACFs occurring at lags of 590 hours and 740 hours, respectively, for the two sets of data.

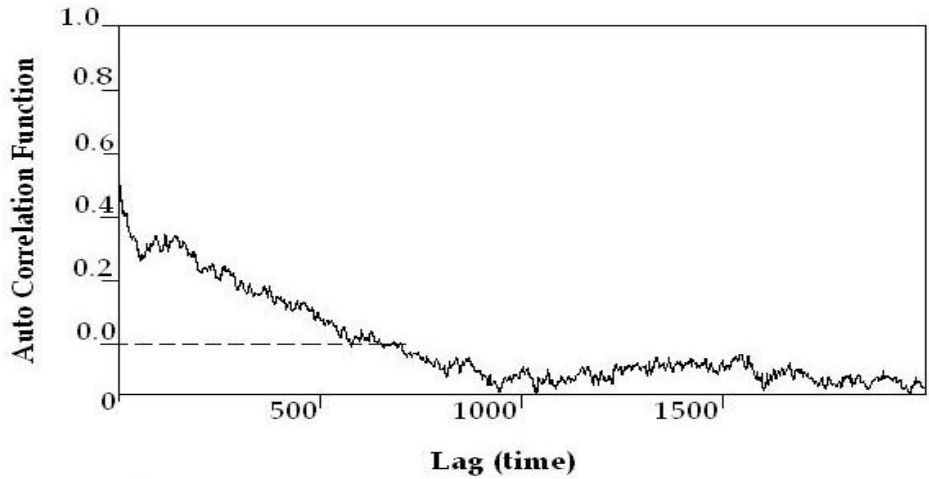

Fig.3a Autocorrelation function of the original helium time series

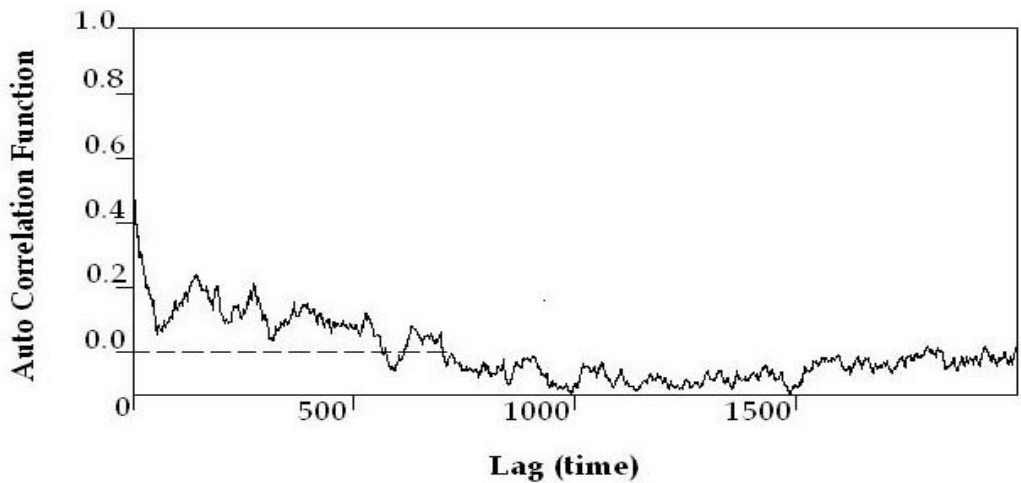

Fig. 3b Autocorrelation function of the filtered helium time series

Figures 4a and 4b show that the power spectral densities, $S(f)$, from the helium time sequences decay exponentially at high frequencies. Both sequences exhibit a power-law decay of the form $S(f) \sim 1/f^{\beta}$, where $\beta = 1.61$ for the smoothed data and $\beta = 1.76$ for the anomalous data.

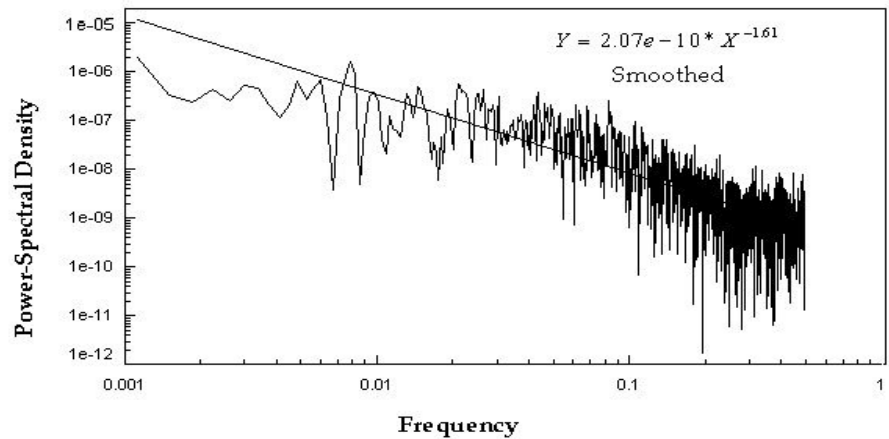

Fig. 4a  Power spectrum of the original helium sequence

These ACFs and power spectra are typical of statistical self-similar processes that have well-defined long-range power-law correlations.

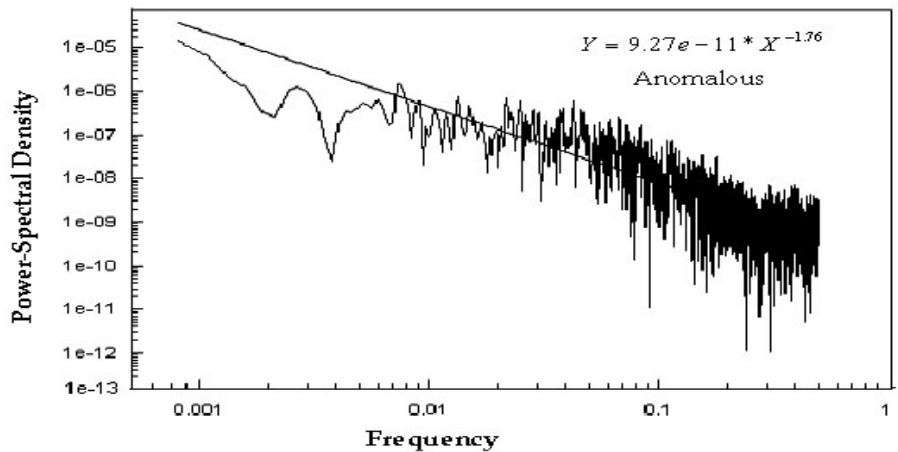

Fig. 4b  Power Spectrum of the filtered helium time sequence

Figure 5 shows the log-log plot of rms fluctuations, $F(m)$, against window size $m$ for the helium sequences over two different temporal phases: A and B. Phase A corresponds to the original anomalous data while phase B corresponds to the smoothed data, after the anomalous peaks were removed.

For comparison, line C in Figure 5 represents a profile for random data generated using the same average and standard deviations as those observed in the original helium sequence. The data in phase A show that the scaling exponent $\alpha$ does not remain constant over the period of measurements. Changes in the correlation exponent over different scales result in a crossover point. Such a crossover, at which scaling behavior changes from one parameter range to another, implies that different mechanisms influence the system dynamics (Peng et al., 1995).

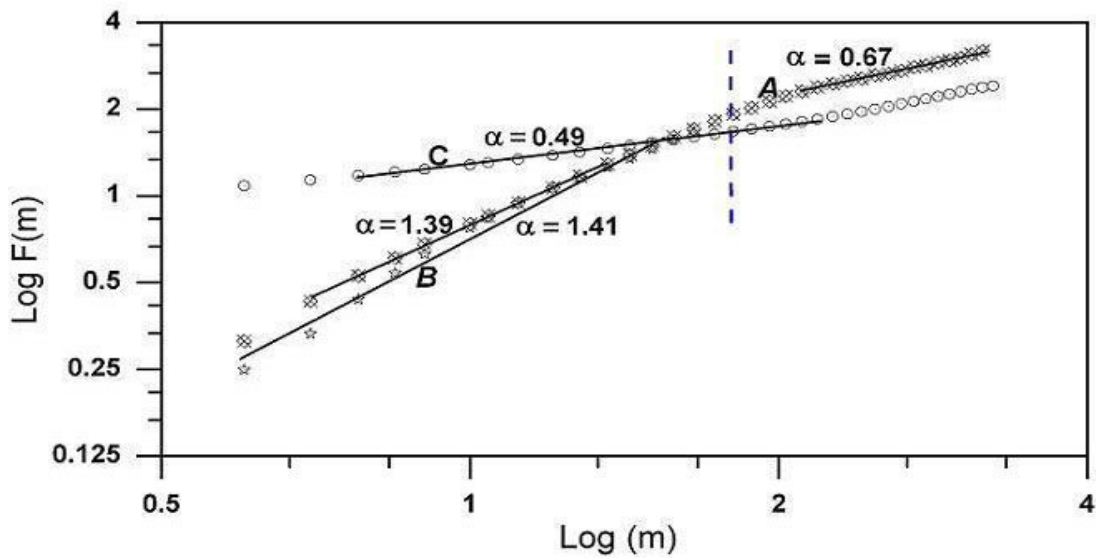

Fig. 5 Plot of log (m) vs. log F(m) derived from helium concentrations revealing two scaling regions.

The DFA approach reveals that the original helium signal contains two scaling exponents that dominate over different time scales. Over short time-scales, the original and smoothed data exhibit traits similar to brown noise ($\alpha \sim 1.39$); however, over longer time-scales, the exponent ($\alpha \sim 0.67$) is typical of persistent long-range correlations. The scaling exponent can also be viewed as an indicator of roughness in the original time series: large values of $\alpha$ correspond to a smoother time series. Therefore, the comparatively low value of $\alpha = 0.67$, which occurs over the larger time scale, originates from increased roughness in the helium sequence. This roughness might be caused by responses to external driving forces, such as seismic perturbations on fluid reservoirs through stress-strain alterations.

Our analysis, therefore, captures a crossover phenomenon and exhibits long-range memory dependence in the helium sequence. It may be that, in the short time-scale regime where the value of $\alpha$ is large, fluctuations in helium concentration are dominated by the earth's vibration. Alternatively, those fluctuations might be attributed to the local geophysical mechanism by which helium escapes from the earth's crust. However, in the longer time-scale regime, helium fluctuations are probably caused by reorganization in the intrinsic dynamics of a complex hydrothermal system, induced by seismic perturbations, which lead to persistent behavior. Precursory signals may well originate in the strain resulting from physical stresses that build up within the crust. Such strain may initiate the earth's early degassing before a seismic tremor. In view of this scenario, abrupt variations in helium signals are thought to be consequences of foreshocks that precede a main seismic event.

## 4. Conclusion

Detrended fluctuation analysis appears to be a credible approach for delineating anomalous signatures in real-time observables; moreover, it gives a plausible quantitative description of correlations in experimental data, even in the presence of nonstationarities or signals having varying time-dependent statistical properties (Hu, et al., 2001). Here, the observed temporal variations in helium concentration may be ascribed to transient responses to an environment that fluctuates over a wide range of time scales. Our results suggest that the emission of helium through fractures and fissures is largely influenced by seismo-tectonic perturbations occurring inside the earth. External seismic stimulation affects the amount of helium released from the earth's interior; such changes in the amounts of released helium are reflected in changes in scaling properties. The occurrence of a crossover point could be a precursor of crustal instability, caused by redistribution of stresses from potential sources of seismic disturbances across the earth's crust.

**Acknowledgements**

The authors are grateful to the Department of Atomic Energy (DAE) and the Department of Science and Technology (DST), Government of India, for sponsoring the activities reported here.